\documentclass{aa}  
\usepackage{graphicx}
\usepackage{txfonts}
\usepackage{lipsum}

\usepackage{orcidlink}

\usepackage{booktabs}
\usepackage{amsmath,amstext}
\usepackage{upgreek}

\usepackage{booktabs}
\usepackage{siunitx}
\usepackage{enumitem} 

\newcommand{\teff}{$T_\mathrm{eff}$}

\newcommand{\logg}{$\log g$}
\newcommand{\feh}{[Fe/H]}
\newcommand{\micro}{$\xi_\mathrm{micro}$}
\newcommand{\cafe}{[Ca/Fe]}
\newcommand{\sife}{[Si/Fe]}
\newcommand{\mgfe}{[Mg/Fe]}

\makeatletter
\@ifpackageloaded{lineno}{%
  \AtBeginDocument{\nolinenumbers}%
}{}

\providecommand{\linenumbers}{}%
\providecommand{\nolinenumbers}{}%
\providecommand{\modulolinenumbers}[1]{}%
\providecommand{\switchlinenumbers}{}%
\providecommand{\runninglinenumbers}{}%
\providecommand{\rightlinenumbers}{}%
\providecommand{\leftlinenumbers}{}%

\@ifundefined{linenumbers}{%
  \newenvironment{linenumbers}{}{}%
}{%
  \renewenvironment{linenumbers}{}{}%
}

\let\thelinenumber\relax
\let\LineNumber\relax
\let\makeLineNumber\relax
\renewcommand\linenumberfont{}%
\makeatother

\begin{document}

\title{Supergiant GCIRS 22 in the Milky Way Nuclear Star Cluster: Revised alpha abundances}

\author{
    B.~Thorsbro\orcidlink{0000-0002-5633-4400}\inst{1,2,3}
\and
    S.~Khalidy\orcidlink{0009-0008-3392-2470}\inst{4}
\and
    R.~M.~Rich\orcidlink{0000-0003-0427-8387}\inst{5}
\and
    M.~Schultheis\orcidlink{0000-0002-6590-1657}\inst{1}
\and
    D.~Taniguchi\orcidlink{0000-0002-2861-4069}\inst{6,7,3}
\and
    A.~M.~Amarsi\orcidlink{0000-0002-3181-3413}\inst{8}
\and
    G.~Kordopatis\orcidlink{0000-0002-9035-3920}\inst{1}
\and
    G.~Nandakumar\orcidlink{0000-0002-6077-2059}\inst{9,2}
\and
    S.~Nishiyama\orcidlink{0000-0002-9440-7172}\inst{10}
\and
    N.~Ryde\orcidlink{0000-0001-6294-3790}\inst{2}
}

\institute{Université Côte d’Azur, Observatoire de la Côte d’Azur, CNRS, Laboratoire Lagrange, 06000 Nice, France\\
\email{brian.thorsbro@oca.eu}
\and
Division of Astrophysics, Department of Physics, Lund University, Box 118, SE-22100 Lund, Sweden
\and
Department of Astronomy, Graduate School of Science, The University of Tokyo, 7-3-1 Hongo, Bunkyo-ku, Tokyo 113-0033, Japan
\and
MAUCA – Master track in Astrophysics, Université Côte d’Azur, Observatoire de la Côte d’Azur, Parc Valrose, 06100 Nice, France
\and
Department of Physics and Astronomy, UCLA, 430 Portola Plaza, Box 951547, Los Angeles, CA 90095-1547, USA
\and
Department of Physics, Tokyo Metropolitan University, 1-1 Minami-Osawa, Hachioji, Tokyo 192-0397, Japan
\and
National Astronomical Observatory of Japan, 2-21-1 Osawa, Mitaka, Tokyo 181-8588, Japan
\and
Theoretical Astrophysics, Department of Physics and Astronomy, Uppsala University, SE-751 20 Uppsala, Sweden
\and
Aryabhatta Research Institute of Observational Sciences, Manora Peak, Nainital 263002, India
\and
Miyagi University of Education, 149 Aramaki-Aza-Aoba, Aoba-ku, Sendai, Miyagi 980-0845, Japan
}

\date{Received September 15, 1996; accepted March 16, 1997}
   
\abstract{ 
The chemical abundances of alpha-elements in Galactic Centre (GC) supergiants provide key insights into the chemical enrichment and star formation history of the Milky Way\'s Nuclear Star Cluster. Previous studies have reported enhanced alpha-element abundances, raising questions about the chemical evolution of this unique region.
}{ 
We aim to reassess the alpha-element abundances in the GC supergiant GCIRS 22 using updated spectral modelling and non-local thermodynamic equilibrium (NLTE) corrections to resolve discrepancies from earlier abundance analyses.
}{ 
High-resolution near-infrared spectra of GCIRS 22 were analysed using contemporary line lists and precise stellar parameters derived from scandium line diagnostics. We applied comprehensive NLTE corrections to accurately determine the abundances of silicon and calcium.
}{ 
Our analysis reveals solar-scale alpha abundances ([Ca/Fe] = 0.06 $\pm$ 0.07; [Si/Fe] = $-$0.08 $\pm$ 0.20) for GCIRS 22, significantly lower than previous LTE-based findings. NLTE corrections reduce the calcium abundance by approximately 0.3\,dex compared to LTE estimates, aligning our results with recent studies and highlighting the importance of accurate NLTE modelling.
}{ 
The solar-scale alpha-element abundances observed in GCIRS 22 suggest that recent star formation in the region has not been dominated by Type II supernovae, such as those expected from a recent starburst. Our findings support a scenario of episodic star formation, characterized by intermittent bursts separated by extended quiescent phases, or potentially driven by gas inflows from the inner disk, funnelled by the Galactic bar. Future comprehensive NLTE studies of additional GC stars will be essential for refining our understanding of the region's chemical evolution and star formation history.
}

\keywords{Stellar abundances (1577) --- Milky Way Galaxy nucleus (565) --- Late-type supergiant stars (910)}

\maketitle

\section{Introduction} \label{sec:intro}

The Galactic Centre (GC) hosts the Milky Way's nuclear star cluster (NSC), a dense, massive stellar system that serves as a unique laboratory for probing star formation, chemical enrichment, and dynamical evolution in extreme environments \citep{neumayer:20, feldmeier-krause:2025}. Alongside the nuclear stellar disk (NSD) and the central molecular zone, the NSC forms a complex ecosystem that holds critical clues to the Milky Way’s formation history and the role of galactic nuclei in galaxy evolution \citep{jwst_nsc_white:23, solanki:2023}. Accurate chemical characterization of its stellar populations is key to unravelling these processes.

The GC’s structure is defined by three primary components: (1) the NSC, a compact, spherical system with a half-light radius of $\sim4$\,pc, diameter of $\sim12$\,pc ($\sim300\arcsec$), and mass of $2\times10^7\,\mathrm{M}_\odot$ \citep{schodel_nsc:2014, gallego-cano:2020}; (2) the NSD, a rotating disk with a break radius of $\sim90$\,pc, scale height of $\sim45$\,pc, and mass of $1\times10^9\,\mathrm{M}_\odot$ \citep{Launhardt:2002, blandhawthorn:16}; and (3) the central molecular zone, a $\sim250\times50$\,pc molecular gas complex \citep{henshaw:2023}. Embedded within the Galactic Bulge/Bar, these structures constitute the Milky Way’s largest reservoir of dense gas, sustained by bar-driven inflows \citep{baba:2020, sormani:2022}.

Star formation in the GC has been episodic, with evidence for both ancient ($8$–$10$\,Gyr) and recent ($<10$\,Myr) populations \citep{matsunaga:2011, NoguerasLara:2020, schodel:2020, thorsbro:2023}. While the NSC and NSD exhibit distinct star formation histories, kinematic and metallicity gradients suggest a smooth transition between them, possibly indicating a shared origin \citep{NoguerasLara:2023}. The NSC’s star formation history remains debated: studies propose either a predominantly old ($>10$\,Gyr) population with minor intermediate-age ($\sim3$\,Gyr) and young ($<100$\,Myr) components \citep{schodel:2020, nogueraslara:2021}, or a dominant $\sim5$\,Gyr population \citep{chen-do:23}. Resolving this requires precise chemical abundances, particularly for young/intermediate-age candidates identified by \citet{nishiyama:16, nishiyama:2023}.

NSCs are ubiquitous in galaxies with masses $\gtrsim10^8$–$10^{10}\,\mathrm{M}_\odot$ and often coexist with supermassive black holes, their masses scaling with host galaxy properties \citep{neumayer:20, kormendy:13, georgiev:2016}. Two primary formation channels have been proposed: (1) globular cluster infall via dynamical friction (timescales $\sim1$\,Gyr; \citep{Tremaine75, Hartmann:2011, Antonini:2012, Tsatsi:2017}) and (2) in situ star formation from accreted gas \citep{Loose82, Nayakshin05}. A hybrid scenario may explain the NSC’s complex population \citep{ArcaSedda2020}, with chemical abundances serving as critical diagnostics.

Recent advancements in high-resolution, near-infrared spectroscopy have enabled detailed chemical analyses despite significant optical extinction toward the Galactic centre. These investigations have demonstrated that stars within the NSC exhibit a broad metallicity range and distinct chemical abundance patterns, reflecting complex star formation histories involving multiple episodes of stellar birth, potentially driven by in situ formation from gas funnelled into the Galactic centre or through infalling stellar clusters \citep{thorsbro:2020, bentley:2022, thorsbro:2023, Nishiyama:2024, nandakumar:2025, ryde:2025, feldmeier-krause:2025}.

The red supergiant population represents the most recent generation of star formation in the Galactic Centre, reflecting what can be assessed as closest to the present-day composition of the interstellar medium. As supergiant stars are cool and luminous, they are in principle amenable to high-resolution abundance analysis, although approaches and techniques have evolved since the pioneering work of \citet{ramirez:00} and \citet{cunha:2007}. Among the most significant findings of \citet{cunha:2007} are striking $+0.3$\,dex enhancements of alpha elements, especially oxygen and calcium, in the supergiant population. Save for one LTE analysis of the Galactic Centre supergiants GCIRS 7 and VR 5--7 \citep{DAvies:2009} the consensus reports elevated abundance ratios in this population, relative to other studies \citep{ryde:2025}; hence the subject is worth a revisit.

Recent work by \citet{thorsbro:2020} has further identified an enhanced silicon abundance at supersolar metallicities within the NSC, highlighting the uniqueness of the alpha-element enrichment processes occurring in this region. In their discussion, \citet{thorsbro:2020} suggest that the alpha-enhanced, metal-rich stars in the NSC could either be young or old, depending on their formation and chemical evolution histories. Profiting from the young age of supergiants, the study of supergiants offers an excellent opportunity to test these scenarios and refine our understanding of the origins of alpha-enhanced populations in the Galactic centre.

Extending abundance analyses past the alpha-elements, \citet{Guerco:2022ApJ} included fluorine in NSC giants, broadening the chemical inventory, and \citet{Guerco:2022MNRAS} showed GCIRS~7 to be fluorine--poor, a signature of CNO--cycle dredge--up, thereby highlighting fluorine as a tracer of internal mixing in massive red supergiants.

The identification of multiple dynamical substructures within the NSC, including distinct stellar disks characterized by varying eccentricities and orbital orientations, further emphasizes the complexity of its formation and evolutionary processes \citep{jia:2023}. The coexistence of young stars and older, metal-poor populations suggests ongoing or episodic star formation activities, possibly associated with transient and dynamic gas inflow mechanisms \citep{thorsbro:2023}. Moreover, unique chemical signatures, such as enhanced sodium abundances, indicate uncommon nucleosynthetic pathways or specific environmental conditions prevalent in the Galactic centre region \citep{ryde:2025}.

Research focusing on individual stars, such as the red supergiant GCIRS\footnote{Named ``InfraRed Source'' (IRS) by \citet{Becklin:1975}. Because SIMBAD now adopts the ``GCIRS'' prefix we follow that convention here.} 22, provides valuable insight into the chemical enrichment and stellar‐evolution processes shaping the NSC.  Together with GCIRS 7 and GCIRS 19, GCIRS 22 is one of only three confirmed red supergiants within the central $\sim$1\,pc of the Galaxy \citep{Blum:1996, ramirez:00, blum2003}.  Its extreme luminosity and relative isolation make it an ideal target for high‐resolution infrared spectroscopy: in this study we analyse Keck II/NIRSPEC observations of GCIRS 22 and compare the derived abundances and stellar parameters with those obtained in earlier studies.
Further, conclusions achieved from the study of GCIRS 22 could potentially be extrapolated to the previous obtained results from the other supergiants studied by \citet{ramirez:00} and \citet{cunha:2007}.

Here, we explore the chemical composition of GCIRS 22 with the benefit of contemporary line lists and improved physical modelling including treatment of non-local thermodynamic equilibrium (NLTE) effects. Details of the observations and data reduction are provided in Section \ref{sec:observations}, followed by details of the analysis in Section \ref{sec:analysis}. In Section \ref{sec:results} we discuss the results and conclude in Section \ref{sec:conclusions}.

\begin{table*}[t]
\caption{Compiled data for GCIRS 22.} 
\label{tab:starsummary}
\centering
\begin{tabular}{lrrrrr}
\toprule
Parameter & C07 & T19 & AP17 & This work & using C07 parameters\\
\midrule
M$_\text{bol}$ [mag]          & $-6.49 \pm 0.42$ &       &         & $-6.27 \pm 0.23$ \\
K$_\text{s}$ [mag]            & 7.39             &       &         & $7.391 \pm 0.024$ \\
Mass [M$_\odot$]              & $10 \pm 2$       &       &         & $7.7 \pm 1.5$ \\
Age [Myr]                     & 25               &       &         & $25$ -- $63$ \\
T$_\text{eff}$ [K]            & $3750 \pm 150$   & 3053  & 3271    & $3350 \pm 150$ \\
$\log g$ [dex]                & $0.2 \pm 0.3$    & 0.24  & 0.30    & $-0.02 \pm 0.15$ \\
$\xi$ [km\,s$^{-1}$]          & $2.3 \pm 0.1$    & 1.76  & 1.71    & $2.53 \pm 0.22$ \\
{[Fe/H]} [dex]                & $0.1 \pm 0.1$    & 0.05  & $-0.05$ & $-0.08 \pm 0.17$ & $0.04 \pm 0.15$ \\
{[Mg/Fe]} [dex]               &                  & 0.18  & $-0.06$ & $-0.07 \pm 0.50$ \\
{[Si/Fe]} [dex]               &                  & 0.17  & $-0.06$ & $-0.08 \pm 0.20$ \\
{[Ca/Fe]} [dex] ``NLTE''      &                  &       &         &  $0.06\pm 0.07$ & $0.26 \pm 0.06$ \\
{[Ca/Fe]} [dex] ``LTE''       & $0.63 \pm 0.1$   & 0.04  &         &  $0.40 \pm 0.08$ & $0.66 \pm 0.06$ \\
\bottomrule
\end{tabular}
\tablefoot{
The coordinates of GCIRS 22 are RA = 17:45:39.83, Dec = $-$29:00:53.90. C07 is data from \citet{cunha:2007}, T19 is data obtained using the PAYNE \citep{ting:2019}, and AP17 is data from APOGEE DR17 \citep{apogee_dr17}. Some uncertainties in T19 and AP17 are reported, but they are not comparable to the uncertainties of C07 and this work, and thus not reported here. In the last column we analysed our spectrum assuming stellar parameters from \citet{cunha:2007}.
}
\end{table*}

\section{Observations} \label{sec:observations}

GCIRS 22 has been observed at medium/high-spectral resolution in the K--band using the NIRSPEC \citep{nirspec_mclean,mclean} facility at Keck~II under the program U095 (PI: Rich), as summarized in Table~\ref{tab:starsummary}. The observations were obtained in the course of a larger program that produced the data for e.g.\ \citet{rich:17,thorsbro:2018,thorsbro:2020,thorsbro:2023}.

We used the $0.432'' \times 12''$ slit and the NIRSPEC-7 filter, giving the resolving power of $R\sim23,000$. We did not use the adaptive optics system to increase photon counts. The stars are observed with an ABBA scheme with a nod throw of $6\arcsec$ along the slit, in order to achieve proper background and dark subtraction. Each exposure was integrating for 250\,s. Five spectral orders are recorded, covering the wavelength range of 21,000--24,000\,\AA. The wavelength coverage is not complete; there are gaps between the orders. The stars are reduced with the NIRSPEC software {\tt redspec} \citep{nirspec_reduction}, and thereafter with IRAF \citep{IRAF} to normalize the continuum, eliminate cosmic-ray hits, and correct for telluric lines. The latter has been done with a high signal-to-noise spectrum of the rapidly rotating O6.5V star HIP 89584. More details about the data reduction are given in \citet{rich:17}.

\section{Analysis}\label{sec:analysis}

Our analysis of the observed star, GCIRS 22, is described in the following subsections. First, the stellar parameters of effective temperature, surface gravity, and microturbulence are derived. This is followed by the line lists used for the spectral analysis and determination of metallicity and abundances. Finally, we employ a Bayesian method grounded in isochrone models to validate the results.

\subsection[Effective temperature]{Effective temperature, \teff}\label{sec:temp}

The effective temperature, \teff, of GCIRS 22 has been determined in previous studies: $3550\pm300$\,K by \citet{ramirez:00}, $3710\pm140$\,K by \citet{blum2003}, $3750\pm150$\,K by \citet{cunha:2007}, 3053\,K by \citet{ting:2019}, and $3271\pm4.7$\,K from the APOGEE DR17 survey \citep{apogee_dr17}. In the studies by \citet{ramirez:00,blum2003,cunha:2007} the temperatures were determined using the equivalent width of the CO bandhead near 2.3\,$\upmu$m. In the study by \citet{ting:2019}, they use a neural network machine learning tool, called PAYNE, and note that the error is on the order of 1\%, i.e. $\pm$30\,K, however, for cool giants like GCIRS 22 they also note that the PAYNE appears to be systemically $\sim$200\,K lower than comparable photometric surveys. In general we choose not to list uncertainties from the PAYNE and APOGEE data as they are not including a treatment of propagated uncertainties leading them to have lower values that are not comparable to the other studies.

We employ the method developed by \citet{thorsbro:2020} to derive \teff\ using the high temperature sensitivity of the scandium lines arising from the 3d$^2$4s – 3d4s4p transition \citep{thorsbro:2018,thorsbro_atoms:20}. The effective temperature determined using this method is $3350\pm150$\,K  and is listed in Table~\ref{tab:starsummary} together with the results from the other studies.

\begin{figure*}
    \centering
    \includegraphics[width=1.0\textwidth]{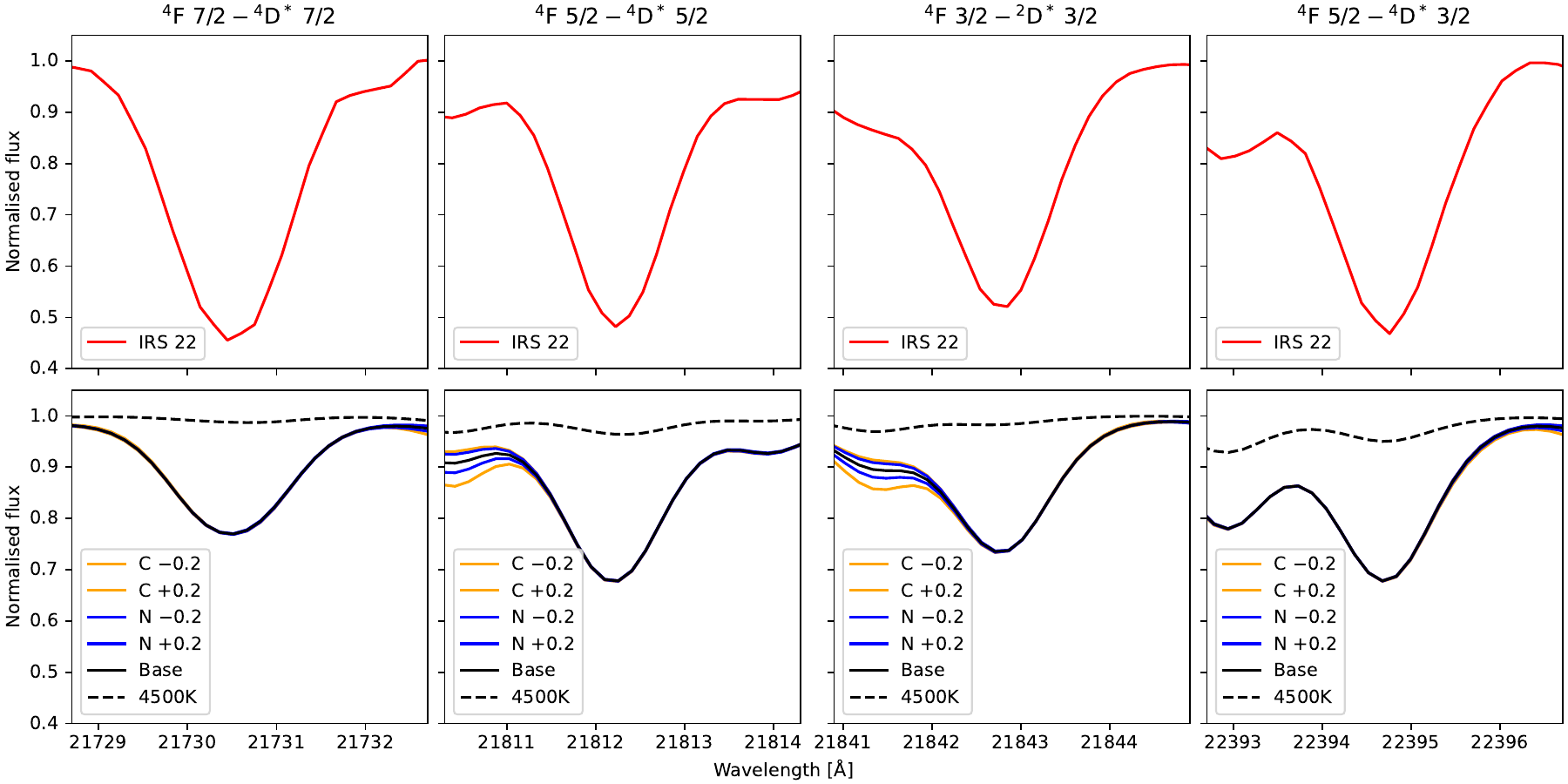}
    \caption{Four fine structure energy level transitions in neutral scandium 3d$^2$4s – 3d4s4p. In the top panel is the observed spectra of GCIRS 22. In the bottom panel: Model spectra showing the sensitivity of the scandium lines to varying the carbon and nitrogen abundances, showing that the scandium lines are not affected by blending of CN molecular line features. The sensitivity analysis was based on a model using an effective temperature of 3600\,K and solar abundances. We have included a model using an effective temperature of 4500\,K to illustrate the temperature sensitivity of the scandium lines as the lines almost vanishes at this temperature due to ionisation.}
    \label{fig:sclinefig}
\end{figure*}

In cool stars, blending of CN molecular lines has to be carefully accounted for when doing spectral analysis. To ensure that the scandium lines used for the temperature determination are not affected by CN blends, we model the spectra and vary the abundances of carbon and nitrogen with $\pm$$0.2$\,dex, to demonstrate that the scandium line features are not affected by CN blends, shown in Figure~\ref{fig:sclinefig}. The line lists used in this modelling are the same as those used for our later analysis and are detailed in Section~\ref{sec:linelist}.

\subsection[Surface gravity]{Surface gravity, \logg}\label{sec:logg}

We use photometry to estimate the surface gravity. Using the K$_\text{s}$-band magnitude of $7.391 \pm 0.024$ from the 2MASS catalogue \citep{2mass}, and an extinction correction of $A_K = 2.107 \pm 0.2$ derived from the colour excess $E(\text{J} - \text{K}_\text{s}) = 3.991$ using the relation $A_K = 0.528 \times E(\text{J} - \text{K}_\text{s})$ \citep{Nishiyama:2009}, the absolute K-band magnitude was computed as K$_{\text{abs}} = -9.31 \pm 0.20$.

A bolometric correction of BC$_{\text{Ks}} = 3.04 \pm 0.14$ \citep{Levesque:2006} was applied to obtain the bolometric magnitude $M_{\text{bol}}=-6.27 \pm 0.23$. This result is in very good agreement with the result of $-6.49\pm0.42$ from \citet{cunha:2007}.

GCIRS 22 is located in the near vicinity of the supermassive black hole. Thus, the distance modulus adopted was $14.59 \pm 0.01$, corresponding to a distance to the supermassive black hole of $8.275\pm0.042$\,kpc \citep{GravityR0:21}. The stellar mass was estimated to be $M = 7.7 \pm 1.5\,\text{M}_\odot$ from comparison with evolutionary tracks \citep{Ekstrom:2012}. With an effective temperature of T$_{\text{eff}} = 3350 \pm 150$\,K, the surface gravity was calculated using the photometric relation:
\begin{multline}
    \log\left(\frac{g}{g_\odot}\right) = \log\left(\frac{M}{M_\odot}\right)  + 4\log\left(\frac{T_\text{eff}}{T_{\text{eff},\odot}}\right)  - 0.4 (M_{\text{Bol},\odot} - M_{\text{Bol},*}),
    \label{eq:logg_determination}
\end{multline}
which yield a value of $\log g = -0.02\pm0.15$. The uncertainty on \logg\ was estimated through standard error propagation from the uncertainties in $T_\text{eff}$, $M_\text{bol}$, and $M$,
\begin{multline}
    \sigma_{\log g} = 
    \left[
    \left( \frac{\sigma_M}{\ln{10} \cdot M}  \right)^2 + 
    \left( \frac{4 \cdot \sigma_{T_\text{eff}}}{\ln{10} \cdot T_\text{eff}} \right)^2 + 
    \left( 0.4 \cdot \sigma_{M_{\text{Bol},*}} \right)^2
    \right]^{1/2}.
    \label{eq:logg_uncertainty}
\end{multline}

We find in the following that the uncertainty of the surface gravity has minimal impact on the results obtained for the metallicity and abundances. Note that the difference in \teff\ between our study and \citet{cunha:2007} translates to about 0.2\,dex difference in \logg, while the remainder comes from the difference in mass.

Using isochrones \citep{Ekstrom:2012} and assuming solar metallicity, which is a reasonable assumption as shown later in Section~\ref{sec:metallicity}, the age of the GCIRS 22 is estimated to be between 25 and 63\,Myr old.

\subsection[Microturbulence]{Microturbulence, \micro}\label{sec:micro}

The microturbulence, $\xi_\mathrm{micro}$, which takes into account the small scale, non-thermal motions in the stellar atmospheres, is important for saturated lines, affecting their line strengths. We estimate this parameter using an empirical relation with surface gravity based on a detailed analysis of spectra of supergiant stars by \citet{taniguchi:2025}. Taking the surface gravity uncertainty of 0.15\,dex into account we thus find $\xi_\mathrm{micro} = 2.53 \pm 0.22$\,km\,s$^{-1}$, which is in agreement with \citet{cunha:2007} that finds $\xi_\mathrm{micro} = 2.3 \pm 0.1$\,km\,s$^{-1}$. In this study the uncertainty of $\xi_\mathrm{micro}$ is not a dominant source of uncertainty for the metallicity and abundance determinations as most lines we use are weak lines low on the curve of growth.

\subsection{Line list}\label{sec:linelist}

In order to synthesize the spectra, an accurate list of atomic and molecular energy level transitions is required. In the list of atomic energy level transitions, we used the solar centre intensity atlas \citep{solar_IR_atlas} to update wavelengths and line strengths (astrophysical $\log gf$-values) \citep{thorsbro:17,Nandakumar:2024}.

Because molecular lines are strong features in our spectra, we include molecular lines in the line list. For CN – which is the most dominant molecule apart from CO, whose lines dominate in the  2.3\,\textmu m bandhead region -- we include the list of \citet{sneden:14}. The CO line data are from \citet{Li:2015}. At the shorter wavelengths of our spectral region SiO, H$_2$O, and OH are important. Line lists for these molecules are included in our line list from, respectively, \citet{langhoff:07,Polyansky:2018,brooke:16}.

To identify clean spectral lines suitable for measurement, we examine the synthetic model spectra of cool M~giants to identify lines that are not blended with molecular features, in a similar manner to the analysis of the scandium lines described above. We have listed the lines used for metallicity and abundance determination in Table~\ref{tab:lines}.
\begin{table}
\caption{Lines used for abundance determinations.}
\label{tab:lines}
\centering
\begin{tabular}{lcr}
\toprule
Wavelength in air [\AA] & $E_\mathrm{exc}$ [eV] & $\log gf$ \\
\midrule
\multicolumn{3}{l}{Mg\,\textsc{I}} \\
21208.009 & 6.7260 & $-0.821$ \\
21225.015 & 6.7265 & $-1.292$ \\
\addlinespace
\multicolumn{3}{l}{Si\,\textsc{I}} \\
21195.298 & 7.2886 & $-0.442$ \\
21819.711 & 6.7206 & $-0.078$ \\
22537.593 & 6.6161 & $-0.296$ \\
\addlinespace
\multicolumn{3}{l}{Ca\,\textsc{I}} \\
22608.009 & 4.6802 & $0.368$ \\
22625.015 & 4.6806 & $0.481$ \\
\addlinespace
\multicolumn{3}{l}{Fe\,\textsc{I}} \\
21284.348 & 3.0714 & $-4.414$ \\
21756.929 & 6.2181 & $-0.562$ \\
21851.422 & 3.6416 & $-3.506$ \\
22473.263 & 6.1189 & \,~\,\,$0.367$ \\
\bottomrule
\end{tabular}
\end{table}

\subsection{Non-LTE corrections}

The spectrum synthesis was carried out without the assumption of local thermodynamic equilibrium (LTE). This was achieved by using pre-computed grids of departure coefficients, $b_{i}=n_{i}/n^{*}_{i}$, where $i$ denotes some level index for NLTE and LTE populations $n$ and $n^{*}$ respectively. The departure coefficients were used to correct the LTE line opacities as described in Section~3 of \citet{sme_code_new}. 

For magnesium, silicon, and calcium, the grids of departure coefficients are those described in \citet{amarsi:grids}. The model atoms come from \citet{amarsi:si} for silicon; and \citet{osorio:mg} and \citet{osorio:ca} for magnesium and calcium, albeit with fine structure collapsed. For iron, the grid of departure coefficients and model atom are those presented in \citet{amarsi:fe_new}.

\subsection{Metallicity \label{sec:metallicity}}

In order to determine metallicities, we synthesize spectra and compare them to the observed spectra. We have chosen to use the spectral synthesis code Spectroscopy Made Easy \citep[SME,][]{sme_code,sme_code_new}, which interpolates on a grid of one-dimensional (1D) MARCS atmosphere models \citep{marcs:08}. These are hydrostatic model atmospheres in spherical geometry, computed assuming LTE, chemical equilibrium, homogeneity, and conservation of the total flux (radiative plus convective, with the convective flux computed using a mixing-length recipe). The SME code has the advantage that it includes a flexible $\chi^2$ minimization tool to find the solution that best fits an observed spectrum in a pre-specified spectral window. It also has a powerful continuum normalization routine. In cool star spectra, extra care is needed to normalize the spectrum in the region of a spectral line under consideration.

\begin{figure}
    \centering
    \includegraphics[width=1.0\columnwidth]{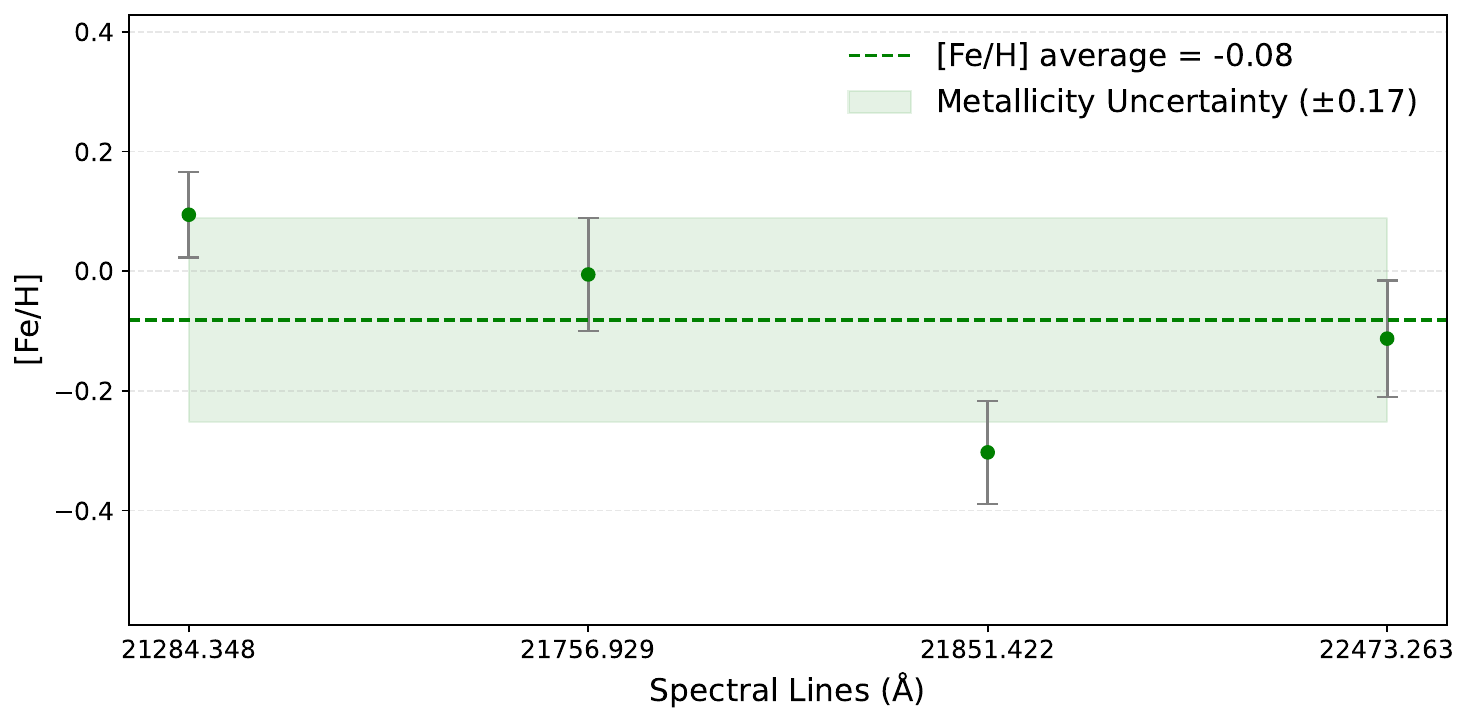}
    \caption{Metallicity determination from individual Fe I Lines. The mean [Fe/H] is found to be $-0.08 \pm 0.17$}
    \label{fig:fehplot} 
\end{figure}

For the determination of metallicity we used the iron lines listed in Table~\ref{tab:lines} and fit them line by line obtaining a metallicity of $-0.08 \pm 0.17$, with the fits shown in Figure~\ref{fig:fehplot}.
To test the robustness of the fit we also fit just iron abundances to the iron lines and the result is $-0.12\pm0.16$, i.e.\ within 0.05, and thus we conclude that the iron lines can be used as basis for the metallicity fit assuming a solar abundance distribution \citep{solar:sme}.
We note that our metallicity determination agrees fairly well with previous studies as shown in Table~\ref{tab:starsummary}.

We included NLTE effects in the modelling of the iron lines. However, the impact is negligible: repeating the analysis in LTE changes the results by only $\sim$0.01\,dex.

\subsection{Abundances} \label{sec:abundances}

\begin{figure}
    \centering
    \includegraphics[width=1.0\columnwidth]{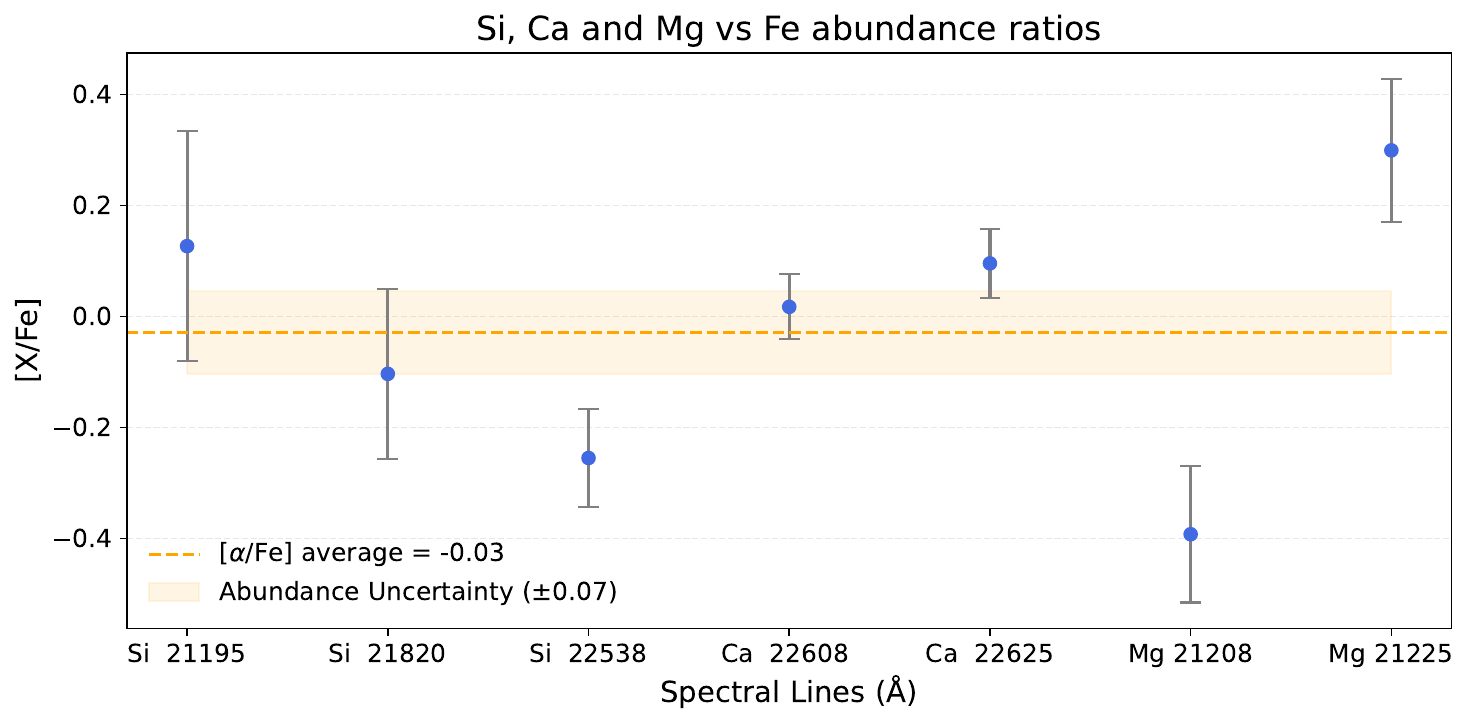}
    \caption{\sife, \cafe\ and \mgfe\ abundance ratios from individual lines. The average abundance of alpha elements was found to be $-0.03 \pm 0.07$. Excluding the magnesium, which can be argued for due to the high scatter, yields and average of $-0.01 \pm 0.09$.}
    \label{fig:alphaplot}
\end{figure}

We determine silicon, calcium and magnesium abundances using the same code and following the same recipe as for determining metallicity, described in Section~\ref{sec:metallicity}. The line by line results for the individual fits are shown in Figure~\ref{fig:alphaplot}

We find \sife\ to be $-0.08 \pm 0.20$, \cafe\ to be $0.06 \pm 0.07$, and \mgfe\ to be $-0.07 \pm 0.50$. Although the magnesium abundances show significant scatter across individual lines we find that whether to include them or not in the average alpha--abundance results is inconsequential. We note that more work is needed on the atomic data for the magnesium lines.

We investigate the impact of NLTE effects on the calcium abundance determination. If we do not include NLTE effects in our synthetic spectrum modelling, we find the determined average calcium abundance value to be \cafe = $0.40\pm0.08$\,dex, i.e.\ $\sim$$0.3$\,dex higher.

We use our spectral analysis method on our observed spectrum assuming the stellar parameters from \citet{cunha:2007} to explore if we can reproduce their calcium abundance value of $0.63\pm0.1$. We find the metallicity to be $0.04 \pm 0.15$, and \cafe\ to be $0.26 \pm 0.06$ with NLTE and $0.66 \pm 0.06$ with LTE. Noting that our analysis very much agree with the previous work, the spectral analysis of the two studies appears to be of the same quality. The difference between the two studies are thus due to the differences in stellar parameters, in particular the effective temperature and mass.

\subsection{Validation of results}

We perform two validation tests of our results using alternative methods. First, we explore the photometric determination of the mass using Bayesian statistics on isochrone models. Second, we explore the spectroscopic analysis using a grid of model spectra and interpolating in the grid to find a global minimum, both described in the following.

We validate the derived stellar age and mass using the Bayesian framework presented in \citet{Kordopatis2023}. This method statistically compares observed stellar parameters against theoretical isochrones while rigorously accounting for measurement uncertainties. The validation process involves projecting observed quantities onto the isochrone grid to compute posterior probability distributions for fundamental parameters, using an age-metallicity prior to constrain the solution space.

For this validation, we employed PARSEC isochrones \citep{Bressan2012} spanning ages from 0.01 to 1\,Gyr in 0.01\,Gyr increments. The Bayesian projection incorporated three key observables: effective temperature ($T_{\text{eff}} = 3350 \pm 150$\,K), bolometric magnitude ($M_{\text{bol}} = -6.27 \pm 0.23$), and metallicity ($\text{[Fe/H]} = -0.08 \pm 0.17$). Surface gravity (log\textit{g}) was excluded as it is derivable from $T_{\text{eff}}$ and $M_{\text{bol}}$ (Eq.~\ref{eq:logg_determination}). A flat age-metallicity prior was adopted to avoid introducing external constraints.

In the posterior distributions fitting with the isochrone models we find \teff\ $=3349\pm87$\,K, $M_{\text{bol}}=-6.14\pm0.40$, \feh\ $=-0.13\pm0.14$, mass $=5.2\pm0.9\,\si{M_\odot}$, \logg\ $=-0.17\pm0.08$, and age $=104\pm30$\,Myr. The derived atmospheric parameters (\teff, \logg, \feh) and luminosity ($M_{\text{bol}}$) are mutually consistent. The two isochrone sets yield masses that differ by only $\sim$$1.4\,\sigma$, which is not statistically significant and thus within uncertainties. The lower--mass solution implies an age $\sim$$60$\,Myr older; nevertheless, both solutions point to a young, massive star. We therefore regard the isochrone--based fit as an independent validation of our preferred solution.

We also validate the spectroscopic analysis of the abundances using the software Starkit \citep{starkit,starkit_2}. It is a method that takes a grid of model spectra and ``interpolates'' in the grid to find a model spectrum that is closest to the observed spectrum using the MulitNest algorithm \citep{multinest}. To construct the grid we calculate models using spectral synthesis code, Spectroscopy Made Easy, described in Section~\ref{sec:metallicity} with the same line lists and model atmospheres. With this method we find the metallicity to be $-0.14 \pm 0.15$ and the alpha--abundances to be $0.16 \pm 0.20$, and conclude that the spectroscopic analysis is also robust.

\section{Results/discussion} \label{sec:results}

Our primary aim is to reassess the alpha-element abundances in Galactic Centre (GC) supergiants through contemporary non-local thermodynamic equilibrium (NLTE) modeling, to clarify the discrepancies observed in previous studies \citep{cunha:2007,thorsbro:2020,ryde:2025}. Using updated spectral modelling, improved line lists, and precise stellar parameters derived from scandium line diagnostics, we find near--solar alpha abundances ([Ca/Fe] = 0.06 $\pm$ 0.07; [Si/Fe] = $-$0.08 $\pm$ 0.20) for GCIRS 22. This result stands in marked contrast to earlier reports of strong alpha-enhancements in GC supergiants, particularly the elevated [Ca/Fe] ratios previously suggested \citep{cunha:2007}.

\begin{figure*}
    \centering
    \includegraphics[width=0.49\textwidth]{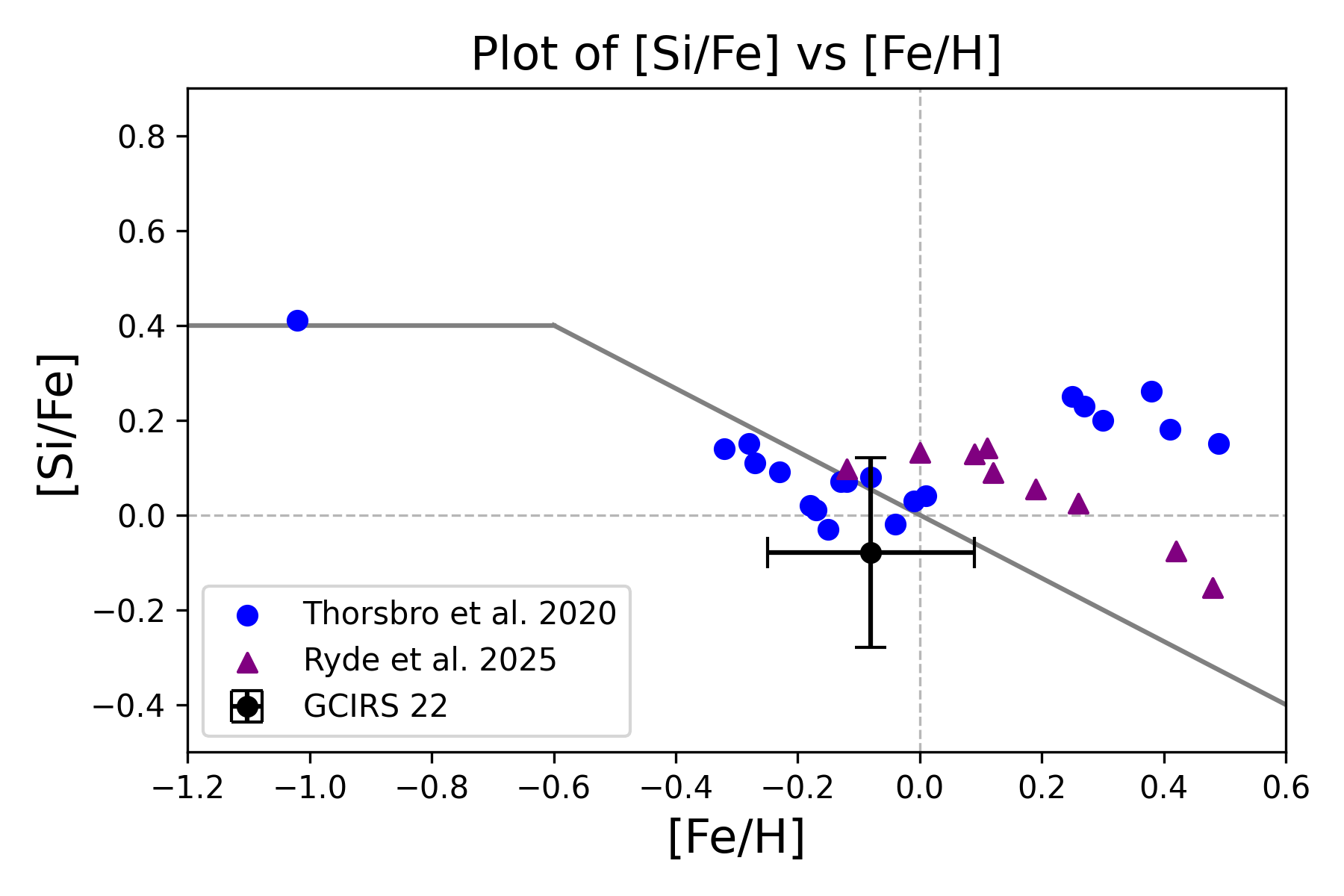}
    \includegraphics[width=0.49\textwidth]{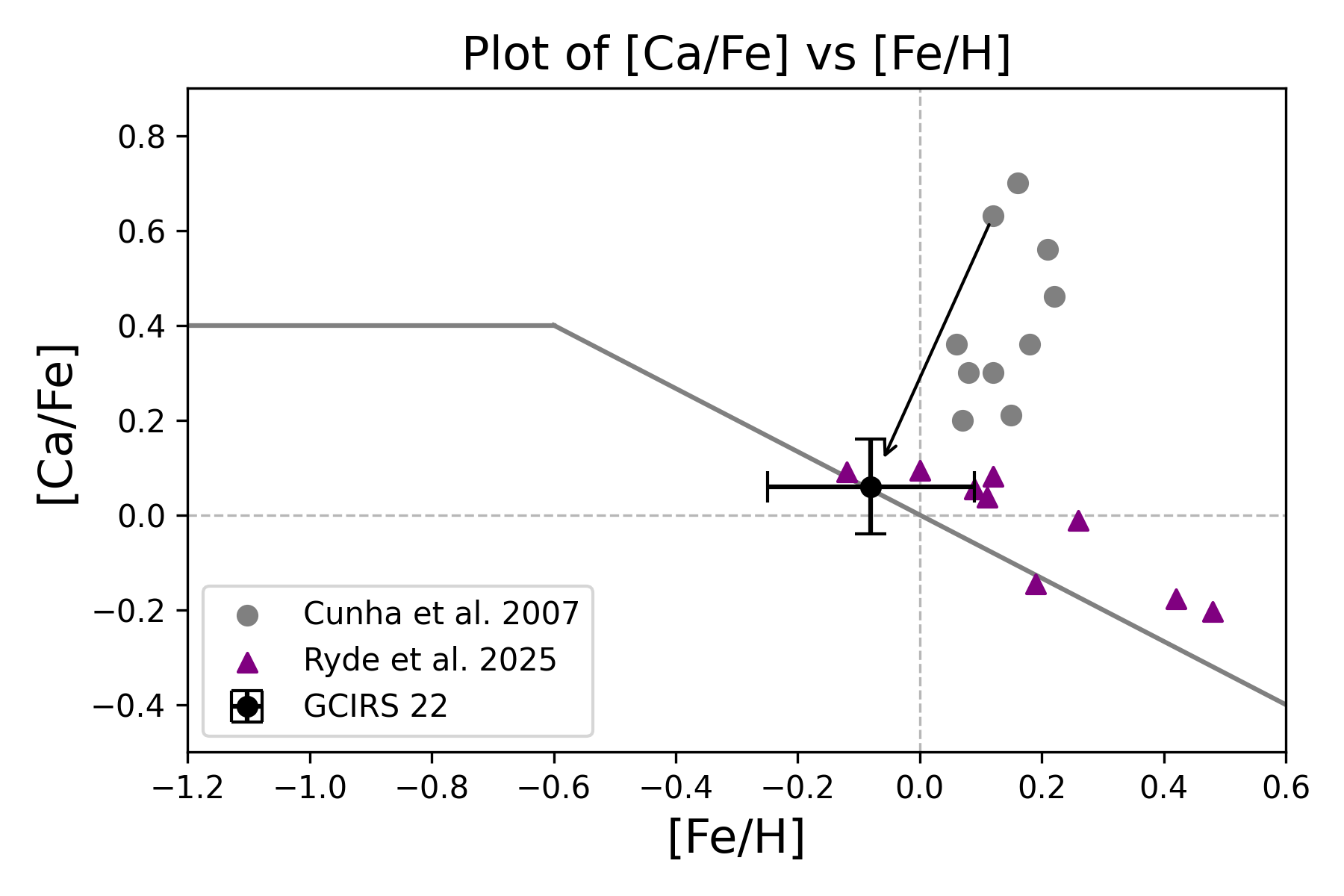}
    \caption{Left panel silicon over iron abundance ratio vs iron abundance, plotting the value obtained for GCIRS 22 and comparing to literature results for the Milky Way Nuclear Star Cluster \citep{thorsbro:2020,ryde:2025}. On the right panel similarly for calcium, however, the gray dots are the results from \citet{cunha:2007} with an arrow indicated the changed value for GCIRS 22. The primary reason for the new value of calcium abundance for GCIRS 22 is due to accounting for NLTE effects, and thus likely to be similar reduction of calcium abundance in the remaining set from \citet{cunha:2007}. The gray line is an ``alpha-knee'' model for a simplified Milky Way Bulge chemical evolution scenario over-plotted for convenience \citep{matteucci:12}.}
    \label{fig:sica}
\end{figure*}

The notable difference between our study and \citet{cunha:2007} primarily stems from the inclusion of NLTE effects. We find that accounting for NLTE reduces the calcium abundance by approximately 0.3 dex compared to LTE analyses, consistent with theoretical predictions by \citet{amarsi:grids}. Figure~\ref{fig:sica} illustrates this adjustment visually, clearly demonstrating how NLTE modelling aligns our findings closer to recent studies reporting solar-scaled alpha abundances \citep{thorsbro:2020,ryde:2025}. It is highly probable that applying similar NLTE corrections to the full \citet{cunha:2007} sample would result in similar reductions, reshaping our understanding of alpha-element enrichment patterns in the GC.

Our findings have significant implications for the chemical evolution scenarios of the Milky Way’s Nuclear Star Cluster (NSC). The solar--scaled alpha abundances we observe suggest that the most recent star formation episodes in the GC, including the formation of supergiants such as GCIRS 22, occurred in an environment already influenced by substantial SN Ia enrichment. Such enrichment would dilute the alpha--enhanced signature typically produced by rapid core--collapse supernovae. The episodic star formation history in the GC, characterized by intermittent bursts followed by extended periods of quiescence, allows sufficient time for SN Ia-driven chemical evolution to dominate the interstellar medium's alpha--element budget \citep{NoguerasLara:2020}. Alternatively, one might expect a more Solar composition for gas if recent star formation is fuelled by flows of gas from the inner disk, channelled by the bar \citep{Athanassoula:1992,Sormani:2020}. Consequently, our derived age for GCIRS 22 (25–63 Myr) aligns well with either scenario, indicating its formation from gas already chemically matured through contributions from both SN Ia and SN II.

Moreover, the observed absence of significant alpha--enhancement in young supergiants such as GCIRS 22 highlights a distinct chemical signature compared to the metal--rich, alpha--enhanced populations previously identified within the NSC by \citet{thorsbro:2020}. \citet{thorsbro:2020} suggests that such alpha-enhanced, metal--rich stars could be either young or old stars. However, we are not able to connect GCIRS 22 to this population, and while we cannot answer the question still, this work would favour an older stellar population (>5 Gyr). This underscores the complexity and heterogeneity of star formation and chemical enrichment processes within the Galactic Centre.

Future investigations incorporating comprehensive NLTE modelling across larger samples of GC stars are essential for fully explaining the chemical evolution of this unique region. Such studies will clarify the interplay between different nucleosynthetic pathways and further refine our understanding of star formation histories in extreme galactic environments.

In summary, our refined analysis clearly indicates solar-scale alpha abundances in recent GC star formation events, significantly revising previous understandings of chemical enrichment patterns in the Galactic Centre and providing crucial constraints for future modeling of galactic chemical evolution.

\section{Conclusions} \label{sec:conclusions}

We present a high--resolution near--infrared spectral analysis of the young red supergiant GCIRS 22 in the Galactic Centre, employing updated atomic data, NLTE corrections, and a refined spectroscopic methodology. Our principal conclusions are:

\begin{enumerate}[itemsep=0.5\baselineskip]
    \item Solar-scale $\upalpha$-abundances: We derive near-solar [$\upalpha$/Fe] ratios for GCIRS 22 ([Ca/Fe] = $0.06 \pm 0.07$; [Si/Fe] = $-0.08 \pm 0.20$). This contrasts with earlier reports of strong $\upalpha$-enhancements in Galactic Centre supergiants \citep{cunha:2007} but aligns with recent studies of the Nuclear Star Cluster \citep{thorsbro:2020,ryde:2025}.

    \item Resolution of historical discrepancies: The 0.57\,dex difference in [Ca/Fe] compared to \citet{cunha:2007} is primarily attributed to NLTE effects (accounting for $\sim$0.34\,dex) and differences in stellar parameters (notably $T_{\rm eff}$ and mass). This underscores the critical importance of NLTE modelling and robust parameter determination for cool supergiants.

    \item Implications for GC chemical evolution: The solar [$\upalpha$/Fe] in GCIRS 22 indicates formation from interstellar medium enriched by both core-collapse SNe and Type Ia SNe.  Yet the presence of a population of supergiants in the NSC along with rich clusters of OB stars (e.g. The Arches cluster) suggests that there might be enough SNe to produce significant alpha enrichment; the Fermi bubbles also point to an energetic and possibly SN powered present.   On the other hand, iron enrichment requires $>100$\,Myr timescales, consistent with episodic star formation in the Galactic Centre \citep{NoguerasLara:2020}. The absence of $\upalpha$-enhancement in this young supergiant is consistent with the $\upalpha$-enhanced, metal-rich populations in the NSC \citep{thorsbro:2020} being ancient relics ($>5$\,Gyr), likely formed during rapid \textit{in situ} star formation.  Given the rich array of current star formation signatures, the earlier era of enrichment may have been witness to a spectacular burst.

\end{enumerate}

Future studies should expand NLTE abundance analyses to larger samples of young supergiants and ancient NSC stars.

\begin{acknowledgements}
BT acknowledges the financial support from the Wenner-Gren Foundation (WGF2022-0041).
SK thanks the Observatoire de la Côte d'Azur and the Université Côte d'Azur for making this project possible.
RMR acknowledges financial support from his late father Jay Baum Rich.
AMA acknowledges support from the Swedish Research Council (VR 2020-03940) and the Crafoord Foundation via the Royal Swedish Academy of Sciences (CR 2024-0015).
GN acknowledges the support from the Crafoord Foundation via the Royal Swedish Academy of Sciences (Vetenskapsakademiens stiftelser; CR 2024-0034).
NR acknowledge support from the Swedish Research Council (grant 2023-04744).
BT and NR acknowledge support from the Royal Physiographic Society in Lund through the Stiftelsen Walter Gyllenbergs and Märta och Erik Holmbergs donations.

\\\indent
The data presented herein were obtained at the W.\ M.\ Keck Observatory, which is operated as a scientific partnership among the California Institute of Technology, the University of California and the National Aeronautics and Space Administration. The Observatory was made possible by the generous financial support of the W.\ M.\ Keck Foundation. The authors wish to recognize and acknowledge the very significant cultural role and reverence that the summit of Mauna Kea has always had within the indigenous Hawaiian community. We are most fortunate to have the opportunity to conduct observations from this mountain.

\\\indent
This publication makes use of data products from the Two Micron All Sky Survey, which is a joint project of the University of Massachusetts and the Infrared Processing and Analysis Center/California Institute of Technology, funded by the National Aeronautics and Space Administration and the National Science Foundation.
\end{acknowledgements}

\bibliographystyle{aa}
\bibliography{references}

\begin{thebibliography}{83}
\expandafter\ifx\csname natexlab\endcsname\relax\def\natexlab#1{#1}\fi

\bibitem[{{Abdurro'uf} {et~al.}(2022){Abdurro'uf}, {Accetta}, {Aerts}, {Silva Aguirre}, {Ahumada}, {Ajgaonkar}, {Filiz Ak}, {Alam}, {Allende Prieto}, {Almeida}, {Anders}, {Anderson}, {Andrews}, {Anguiano}, {Aquino-Ort{\'\i}z}, {Arag{\'o}n-Salamanca}, {Argudo-Fern{\'a}ndez}, {Ata}, {Aubert}, {Avila-Reese}, {Badenes}, {Barb{\'a}}, {Barger}, {Barrera-Ballesteros}, {Beaton}, {Beers}, {Belfiore}, {Bender}, {Bernardi}, {Bershady}, {Beutler}, {Bidin}, {Bird}, {Bizyaev}, {Blanc}, {Blanton}, {Boardman}, {Bolton}, {Boquien}, {Borissova}, {Bovy}, {Brandt}, {Brown}, {Brownstein}, {Brusa}, {Buchner}, {Bundy}, {Burchett}, {Bureau}, {Burgasser}, {Cabang}, {Campbell}, {Cappellari}, {Carlberg}, {Wanderley}, {Carrera}, {Cash}, {Chen}, {Chen}, {Cherinka}, {Chiappini}, {Choi}, {Chojnowski}, {Chung}, {Clerc}, {Cohen}, {Comerford}, {Comparat}, {da Costa}, {Covey}, {Crane}, {Cruz-Gonzalez}, {Culhane}, {Cunha}, {Dai}, {Damke}, {Darling}, {Davidson}, {Davies}, {Dawson}, {De Lee}, {Diamond-Stanic}, {Cano-D{\'\i}az}, {S{\'a}nchez},
  {Donor}, {Duckworth}, {Dwelly}, {Eisenstein}, {Elsworth}, {Emsellem}, {Eracleous}, {Escoffier}, {Fan}, {Farr}, {Feng}, {Fern{\'a}ndez-Trincado}, {Feuillet}, {Filipp}, {Fillingham}, {Frinchaboy}, {Fromenteau}, {Galbany}, {Garc{\'\i}a}, {Garc{\'\i}a-Hern{\'a}ndez}, {Ge}, {Geisler}, {Gelfand}, {G{\'e}ron}, {Gibson}, {Goddy}, {Godoy-Rivera}, {Grabowski}, {Green}, {Greener}, {Grier}, {Griffith}, {Guo}, {Guy}, {Hadjara}, {Harding}, {Hasselquist}, {Hayes}, {Hearty}, {Hern{\'a}ndez}, {Hill}, {Hogg}, {Holtzman}, {Horta}, {Hsieh}, {Hsu}, {Hsu}, {Huber}, {Huertas-Company}, {Hutchinson}, {Hwang}, {Ibarra-Medel}, {Chitham}, {Ilha}, {Imig}, {Jaekle}, {Jayasinghe}, {Ji}, {Johnson}, {Jones}, {J{\"o}nsson}, {Katkov}, {Khalatyan}, {Kinemuchi}, {Kisku}, {Knapen}, {Kneib}, {Kollmeier}, {Kong}, {Kounkel}, {Kreckel}, {Krishnarao}, {Lacerna}, {Lane}, {Langgin}, {Lavender}, {Law}, {Lazarz}, {Leung}, {Leung}, {Lewis}, {Li}, {Li}, {Lian}, {Liang}, {Lin}, {Lin}, {Lin}, {Lintott}, {Long}, {Longa-Pe{\~n}a}, {L{\'o}pez-Cob{\'a}}, {Lu},
  {Lundgren}, {Luo}, {Mackereth}, {de la Macorra}, {Mahadevan}, {Majewski}, {Manchado}, {Mandeville}, {Maraston}, {Margalef-Bentabol}, {Masseron}, {Masters}, {Mathur}, {McDermid}, {Mckay}, {Merloni}, {Merrifield}, {Meszaros}, {Miglio}, {Di Mille}, {Minniti}, {Minsley}, \& {Monachesi}}]{apogee_dr17}
{Abdurro'uf}, {Accetta}, K., {Aerts}, C., {et~al.} 2022, \apjs, 259, 35

\bibitem[{{Amarsi} \& {Asplund}(2017)}]{amarsi:si}
{Amarsi}, A.~M. \& {Asplund}, M. 2017, \mnras, 464, 264

\bibitem[{{Amarsi} {et~al.}(2022){Amarsi}, {Liljegren}, \& {Nissen}}]{amarsi:fe_new}
{Amarsi}, A.~M., {Liljegren}, S., \& {Nissen}, P.~E. 2022, \aap, 668, A68

\bibitem[{{Amarsi} {et~al.}(2020){Amarsi}, {Lind}, {Osorio}, {Nordlander}, {Bergemann}, {Reggiani}, {Wang}, {Buder}, {Asplund}, {Barklem}, {Wehrhahn}, {Sk{\'u}lad{\'o}ttir}, {Kobayashi}, {Karakas}, {Gao}, {Bland-Hawthorn}, {de Silva}, {Kos}, {Lewis}, {Martell}, {Sharma}, {Simpson}, {Zucker}, {{\v{C}}otar}, {Horner}, \& {GALAH Collaboration}}]{amarsi:grids}
{Amarsi}, A.~M., {Lind}, K., {Osorio}, Y., {et~al.} 2020, \aap, 642, A62

\bibitem[{{Antonini} {et~al.}(2012){Antonini}, {Capuzzo-Dolcetta}, {Mastrobuono-Battisti}, \& {Merritt}}]{Antonini:2012}
{Antonini}, F., {Capuzzo-Dolcetta}, R., {Mastrobuono-Battisti}, A., \& {Merritt}, D. 2012, \apj, 750, 111

\bibitem[{{Arca Sedda} {et~al.}(2020){Arca Sedda}, {Gualandris}, {Do}, {Feldmeier-Krause}, {Neumayer}, \& {Erkal}}]{ArcaSedda2020}
{Arca Sedda}, M., {Gualandris}, A., {Do}, T., {et~al.} 2020, \apjl, 901, L29

\bibitem[{{Athanassoula}(1992)}]{Athanassoula:1992}
{Athanassoula}, E. 1992, \mnras, 259, 345

\bibitem[{{Baba} \& {Kawata}(2020)}]{baba:2020}
{Baba}, J. \& {Kawata}, D. 2020, \mnras, 492, 4500

\bibitem[{{Becklin} \& {Neugebauer}(1975)}]{Becklin:1975}
{Becklin}, E.~E. \& {Neugebauer}, G. 1975, \apjl, 200, L71

\bibitem[{{Bentley} {et~al.}(2022){Bentley}, {Do}, {Kerzendorf}, {Chu}, {Chen}, {Konopacky}, \& {Ghez}}]{bentley:2022}
{Bentley}, R.~O., {Do}, T., {Kerzendorf}, W., {et~al.} 2022, \apj, 925, 77

\bibitem[{{Bland-Hawthorn} \& {Gerhard}(2016)}]{blandhawthorn:16}
{Bland-Hawthorn}, J. \& {Gerhard}, O. 2016, \araa, 54, 529

\bibitem[{{Blum} {et~al.}(2003){Blum}, {Ram{\'{\i}}rez}, {Sellgren}, \& {Olsen}}]{blum2003}
{Blum}, R.~D., {Ram{\'{\i}}rez}, S.~V., {Sellgren}, K., \& {Olsen}, K. 2003, \apj, 597, 323

\bibitem[{{Blum} {et~al.}(1996){Blum}, {Sellgren}, \& {Depoy}}]{Blum:1996}
{Blum}, R.~D., {Sellgren}, K., \& {Depoy}, D.~L. 1996, \aj, 112, 1988

\bibitem[{{Bressan} {et~al.}(2012){Bressan}, {Marigo}, {Girardi}, {Salasnich}, {Dal Cero}, {Rubele}, \& {Nanni}}]{Bressan2012}
{Bressan}, A., {Marigo}, P., {Girardi}, L., {et~al.} 2012, \mnras, 427, 127

\bibitem[{{Brooke} {et~al.}(2016){Brooke}, {Bernath}, {Western}, {Sneden}, {Af{\textcommabelow s}ar}, {Li}, \& {Gordon}}]{brooke:16}
{Brooke}, J. S.~A., {Bernath}, P.~F., {Western}, C.~M., {et~al.} 2016, \jqsrt, 168, 142

\bibitem[{{Chen} {et~al.}(2023){Chen}, {Do}, {Ghez}, {Hosek}, {Feldmeier-Krause}, {Chu}, {Bentley}, {Lu}, \& {Morris}}]{chen-do:23}
{Chen}, Z., {Do}, T., {Ghez}, A.~M., {et~al.} 2023, \apj, 944, 79

\bibitem[{{Cunha} {et~al.}(2007){Cunha}, {Sellgren}, {Smith}, {et~al.}}]{cunha:2007}
{Cunha}, K., {Sellgren}, K., {Smith}, V.~V., {et~al.} 2007, \apj, 669, 1011

\bibitem[{{Davies} {et~al.}(2009){Davies}, {Origlia}, {Kudritzki}, {Figer}, {Rich}, \& {Najarro}}]{DAvies:2009}
{Davies}, B., {Origlia}, L., {Kudritzki}, R.-P., {et~al.} 2009, \apj, 694, 46

\bibitem[{{Do} {et~al.}(2015){Do}, {Kerzendorf}, {Winsor}, {St{\o}stad}, {Morris}, {Lu}, \& {Ghez}}]{starkit_2}
{Do}, T., {Kerzendorf}, W., {Winsor}, N., {et~al.} 2015, \apj, 809, 143

\bibitem[{{Ekstr{\"o}m} {et~al.}(2012){Ekstr{\"o}m}, {Georgy}, {Eggenberger}, {Meynet}, {Mowlavi}, {Wyttenbach}, {Granada}, {Decressin}, {Hirschi}, {Frischknecht}, {Charbonnel}, \& {Maeder}}]{Ekstrom:2012}
{Ekstr{\"o}m}, S., {Georgy}, C., {Eggenberger}, P., {et~al.} 2012, \aap, 537, A146

\bibitem[{{Feldmeier-Krause} {et~al.}(2025){Feldmeier-Krause}, {Neumayer}, {Seth}, {van de Ven}, {Hilker}, {Kissler-Patig}, {Kuntschner}, {L{\"u}tzgendorf}, {Mastrobuono-Battisti}, {Nogueras-Lara}, {Perets}, {Sch{\"o}del}, \& {Zocchi}}]{feldmeier-krause:2025}
{Feldmeier-Krause}, A., {Neumayer}, N., {Seth}, A., {et~al.} 2025, \aap, 696, A213

\bibitem[{{Feroz} \& {Hobson}(2008)}]{multinest}
{Feroz}, F. \& {Hobson}, M.~P. 2008, \mnras, 384, 449

\bibitem[{{Gallego-Cano} {et~al.}(2020){Gallego-Cano}, {Sch{\"o}del}, {Nogueras-Lara}, {Dong}, {Shahzamanian}, {Fritz}, {Gallego-Calvente}, \& {Neumayer}}]{gallego-cano:2020}
{Gallego-Cano}, E., {Sch{\"o}del}, R., {Nogueras-Lara}, F., {et~al.} 2020, \aap, 634, A71

\bibitem[{{Georgiev} {et~al.}(2016){Georgiev}, {B{\"o}ker}, {Leigh}, {L{\"u}tzgendorf}, \& {Neumayer}}]{georgiev:2016}
{Georgiev}, I.~Y., {B{\"o}ker}, T., {Leigh}, N., {L{\"u}tzgendorf}, N., \& {Neumayer}, N. 2016, \mnras, 457, 2122

\bibitem[{{GRAVITY Collaboration} {et~al.}(2021){GRAVITY Collaboration}, {Abuter}, {Amorim}, {Baub{\"o}ck}, {Berger}, {Bonnet}, {Brandner}, {Cl{\'e}net}, {Davies}, {de Zeeuw}, {Dexter}, {Dallilar}, {Drescher}, {Eckart}, {Eisenhauer}, {F{\"o}rster Schreiber}, {Garcia}, {Gao}, {Gendron}, {Genzel}, {Gillessen}, {Habibi}, {Haubois}, {Hei{\ss}el}, {Henning}, {Hippler}, {Horrobin}, {Jim{\'e}nez-Rosales}, {Jochum}, {Jocou}, {Kaufer}, {Kervella}, {Lacour}, {Lapeyr{\`e}re}, {Le Bouquin}, {L{\'e}na}, {Lutz}, {Nowak}, {Ott}, {Paumard}, {Perraut}, {Perrin}, {Pfuhl}, {Rabien}, {Rodr{\'\i}guez-Coira}, {Shangguan}, {Shimizu}, {Scheithauer}, {Stadler}, {Straub}, {Straubmeier}, {Sturm}, {Tacconi}, {Vincent}, {von Fellenberg}, {Waisberg}, {Widmann}, {Wieprecht}, {Wiezorrek}, {Woillez}, {Yazici}, {Young}, \& {Zins}}]{GravityR0:21}
{GRAVITY Collaboration}, {Abuter}, R., {Amorim}, A., {et~al.} 2021, \aap, 647, A59

\bibitem[{{Grevesse} {et~al.}(2007){Grevesse}, {Asplund}, \& {Sauval}}]{solar:sme}
{Grevesse}, N., {Asplund}, M., \& {Sauval}, A.~J. 2007, \ssr, 130, 105

\bibitem[{{Guer{\c{c}}o} {et~al.}(2022{\natexlab{a}}){Guer{\c{c}}o}, {Ram{\'\i}rez}, {Cunha}, {Smith}, {Prantzos}, {Sellgren}, \& {Daflon}}]{Guerco:2022ApJ}
{Guer{\c{c}}o}, R., {Ram{\'\i}rez}, S., {Cunha}, K., {et~al.} 2022{\natexlab{a}}, \apj, 929, 24

\bibitem[{{Guer{\c{c}}o} {et~al.}(2022{\natexlab{b}}){Guer{\c{c}}o}, {Smith}, {Cunha}, {Ekstr{\"o}m}, {Abia}, {Plez}, {Meynet}, {Ramirez}, {Prantzos}, {Sellgren}, {Hayes}, \& {Majewski}}]{Guerco:2022MNRAS}
{Guer{\c{c}}o}, R., {Smith}, V.~V., {Cunha}, K., {et~al.} 2022{\natexlab{b}}, \mnras, 516, 2801

\bibitem[{{Gustafsson} {et~al.}(2008){Gustafsson}, {Edvardsson}, {Eriksson}, {et~al.}}]{marcs:08}
{Gustafsson}, B., {Edvardsson}, B., {Eriksson}, K., {et~al.} 2008, \aap, 486, 951

\bibitem[{{Hartmann} {et~al.}(2011){Hartmann}, {Debattista}, {Seth}, {Cappellari}, \& {Quinn}}]{Hartmann:2011}
{Hartmann}, M., {Debattista}, V.~P., {Seth}, A., {Cappellari}, M., \& {Quinn}, T.~R. 2011, \mnras, 418, 2697

\bibitem[{{Henshaw} {et~al.}(2023){Henshaw}, {Barnes}, {Battersby}, {Ginsburg}, {Sormani}, \& {Walker}}]{henshaw:2023}
{Henshaw}, J.~D., {Barnes}, A.~T., {Battersby}, C., {et~al.} 2023, in Astronomical Society of the Pacific Conference Series, Vol. 534, Protostars and Planets VII, ed. S.~{Inutsuka}, Y.~{Aikawa}, T.~{Muto}, K.~{Tomida}, \& M.~{Tamura}, 83

\bibitem[{{Jia} {et~al.}(2023){Jia}, {Xu}, {Lu}, {Chu}, {O'Neil}, {Drechsler}, {Hosek}, {Sakai}, {Do}, {Ciurlo}, {Gautam}, {Ghez}, {Becklin}, {Morris}, \& {Bentley}}]{jia:2023}
{Jia}, S., {Xu}, N., {Lu}, J.~R., {et~al.} 2023, \apj, 949, 18

\bibitem[{Kerzendorf \& Do(2015)}]{starkit}
Kerzendorf, W. \& Do, T. 2015, Starkit: second release, zenodo, doi:10.5281/zenodo.1117920

\bibitem[{{Kim} {et~al.}(2015){Kim}, {Prato}, \& {McLean}}]{nirspec_reduction}
{Kim}, S., {Prato}, L., \& {McLean}, I. 2015, {REDSPEC: NIRSPEC data reduction}, Astrophysics Source Code Library

\bibitem[{{Kordopatis} {et~al.}(2023){Kordopatis}, {Schultheis}, {McMillan}, {Palicio}, {de Laverny}, {Recio-Blanco}, {Creevey}, {{\'A}lvarez}, {Andrae}, {Poggio}, {Spitoni}, {Contursi}, {Zhao}, {Oreshina-Slezak}, {Ordenovic}, \& {Bijaoui}}]{Kordopatis2023}
{Kordopatis}, G., {Schultheis}, M., {McMillan}, P.~J., {et~al.} 2023, \aap, 669, A104

\bibitem[{{Kormendy} \& {Ho}(2013)}]{kormendy:13}
{Kormendy}, J. \& {Ho}, L.~C. 2013, \araa, 51, 511

\bibitem[{{Langhoff} \& {Bauschlicher}(1993)}]{langhoff:07}
{Langhoff}, S.~R. \& {Bauschlicher}, C.~W. 1993, Chem. Phys. Letters, 211, 305

\bibitem[{{Launhardt} {et~al.}(2002){Launhardt}, {Zylka}, \& {Mezger}}]{Launhardt:2002}
{Launhardt}, R., {Zylka}, R., \& {Mezger}, P.~G. 2002, \aap, 384, 112

\bibitem[{{Levesque} {et~al.}(2006){Levesque}, {Massey}, {Olsen}, {Plez}, {Meynet}, \& {Maeder}}]{Levesque:2006}
{Levesque}, E.~M., {Massey}, P., {Olsen}, K.~A.~G., {et~al.} 2006, \apj, 645, 1102

\bibitem[{{Li} {et~al.}(2015){Li}, {Gordon}, {Rothman}, {Tan}, {Hu}, {Kassi}, {Campargue}, \& {Medvedev}}]{Li:2015}
{Li}, G., {Gordon}, I.~E., {Rothman}, L.~S., {et~al.} 2015, \apjs, 216, 15

\bibitem[{{Livingston} \& {Wallace}(1991)}]{solar_IR_atlas}
{Livingston}, W. \& {Wallace}, L. 1991, {An atlas of the solar spectrum in the infrared from 1850 to 9000 cm-1 (1.1 to 5.4 micrometer)} (NSO Technical Report, Tucson: National Solar Observatory, National Optical Astronomy Observatory, 1991)

\bibitem[{{Loose} {et~al.}(1982){Loose}, {Kruegel}, \& {Tutukov}}]{Loose82}
{Loose}, H.~H., {Kruegel}, E., \& {Tutukov}, A. 1982, \aap, 105, 342

\bibitem[{{Matsunaga} {et~al.}(2011){Matsunaga}, {Kawadu}, {Nishiyama}, {Nagayama}, {Kobayashi}, {Tamura}, {Bono}, {Feast}, \& {Nagata}}]{matsunaga:2011}
{Matsunaga}, N., {Kawadu}, T., {Nishiyama}, S., {et~al.} 2011, \nat, 477, 188

\bibitem[{{Matteucci}(2012)}]{matteucci:12}
{Matteucci}, F. 2012, {Chemical Evolution of Galaxies} (Springer-Verlag Berlin Heidelberg)

\bibitem[{{McLean}(2005)}]{nirspec_mclean}
{McLean}, I.~S. 2005, in High Resolution Infrared Spectroscopy in Astronomy, ed. H.~U. {K{\"a}ufl}, R.~{Siebenmorgen}, \& A.~F.~M. {Moorwood} (Springer Berlin, Heidelberg), 25

\bibitem[{{McLean} {et~al.}(2007){McLean}, {Prato}, {McGovern}, {et~al.}}]{mclean}
{McLean}, I.~S., {Prato}, L., {McGovern}, M.~R., {et~al.} 2007, \apj, 658, 1217

\bibitem[{{Nandakumar} {et~al.}(2024){Nandakumar}, {Ryde}, {Forsberg}, {Montelius}, {Mace}, {J{\"o}nsson}, \& {Thorsbro}}]{Nandakumar:2024}
{Nandakumar}, G., {Ryde}, N., {Forsberg}, R., {et~al.} 2024, \aap, 684, A15

\bibitem[{{Nandakumar} {et~al.}(2025){Nandakumar}, {Ryde}, {Schultheis}, {Rich}, {di Matteo}, {Thorsbro}, \& {Mace}}]{nandakumar:2025}
{Nandakumar}, G., {Ryde}, N., {Schultheis}, M., {et~al.} 2025, \apjl, 982, L14

\bibitem[{{Nayakshin} \& {Cuadra}(2005)}]{Nayakshin05}
{Nayakshin}, S. \& {Cuadra}, J. 2005, \aap, 437, 437

\bibitem[{{Neumayer} {et~al.}(2020){Neumayer}, {Seth}, \& {B{\"o}ker}}]{neumayer:20}
{Neumayer}, N., {Seth}, A., \& {B{\"o}ker}, T. 2020, {A\&ARv}, 28, 4

\bibitem[{{Nishiyama} {et~al.}(2023){Nishiyama}, {Funamoto}, \& {Sch{\"o}del}}]{nishiyama:2023}
{Nishiyama}, S., {Funamoto}, N., \& {Sch{\"o}del}, R. 2023, \apj, 951, 148

\bibitem[{{Nishiyama} {et~al.}(2024){Nishiyama}, {Kara}, {Thorsbro}, {Saida}, {Takamori}, {Takahashi}, {Ohgami}, {Ichikawa}, \& {Sch{\"o}del}}]{Nishiyama:2024}
{Nishiyama}, S., {Kara}, T., {Thorsbro}, B., {et~al.} 2024, Proceedings of the Japan Academy, Series B, 100, 86

\bibitem[{{Nishiyama} {et~al.}(2016){Nishiyama}, {Sch{\"o}del}, {Yoshikawa}, {Nagata}, {Minowa}, \& {Tamura}}]{nishiyama:16}
{Nishiyama}, S., {Sch{\"o}del}, R., {Yoshikawa}, T., {et~al.} 2016, \aap, 588, A49

\bibitem[{{Nishiyama} {et~al.}(2009){Nishiyama}, {Tamura}, {Hatano}, {Kato}, {Tanab{\'e}}, {Sugitani}, \& {Nagata}}]{Nishiyama:2009}
{Nishiyama}, S., {Tamura}, M., {Hatano}, H., {et~al.} 2009, \apj, 696, 1407

\bibitem[{{Nogueras-Lara} {et~al.}(2023){Nogueras-Lara}, {Feldmeier-Krause}, {Sch{\"o}del}, {Sormani}, {de Lorenzo-C{\'a}ceres}, {Mastrobuono-Battisti}, {Schultheis}, {Neumayer}, {Rich}, \& {Nieuwmunster}}]{NoguerasLara:2023}
{Nogueras-Lara}, F., {Feldmeier-Krause}, A., {Sch{\"o}del}, R., {et~al.} 2023, \aap, 680, A75

\bibitem[{{Nogueras-Lara} {et~al.}(2020){Nogueras-Lara}, {Sch{\"o}del}, {Gallego-Calvente}, {Gallego-Cano}, {Shahzamanian}, {Dong}, {Neumayer}, {Hilker}, {Najarro}, {Nishiyama}, {Feldmeier-Krause}, {Girard}, \& {Cassisi}}]{NoguerasLara:2020}
{Nogueras-Lara}, F., {Sch{\"o}del}, R., {Gallego-Calvente}, A.~T., {et~al.} 2020, Nature Astronomy, 4, 377

\bibitem[{{Nogueras-Lara} {et~al.}(2021){Nogueras-Lara}, {Sch{\"o}del}, \& {Neumayer}}]{nogueraslara:2021}
{Nogueras-Lara}, F., {Sch{\"o}del}, R., \& {Neumayer}, N. 2021, \apj, 920, 97

\bibitem[{{Osorio} \& {Barklem}(2016)}]{osorio:mg}
{Osorio}, Y. \& {Barklem}, P.~S. 2016, \aap, 586, A120

\bibitem[{{Osorio} {et~al.}(2019){Osorio}, {Lind}, {Barklem}, {Allende Prieto}, \& {Zatsarinny}}]{osorio:ca}
{Osorio}, Y., {Lind}, K., {Barklem}, P.~S., {Allende Prieto}, C., \& {Zatsarinny}, O. 2019, \aap, 623, A103

\bibitem[{{Piskunov} \& {Valenti}(2017)}]{sme_code_new}
{Piskunov}, N. \& {Valenti}, J.~A. 2017, \aap, 597, A16

\bibitem[{{Polyansky} {et~al.}(2018){Polyansky}, {Kyuberis}, {Zobov}, {Tennyson}, {Yurchenko}, \& {Lodi}}]{Polyansky:2018}
{Polyansky}, O.~L., {Kyuberis}, A.~A., {Zobov}, N.~F., {et~al.} 2018, \mnras, 480, 2597

\bibitem[{{Ram{\'{\i}}rez} {et~al.}(2000){Ram{\'{\i}}rez}, {Stephens}, {Frogel}, \& {DePoy}}]{ramirez:00}
{Ram{\'{\i}}rez}, S.~V., {Stephens}, A.~W., {Frogel}, J.~A., \& {DePoy}, D.~L. 2000, \aj, 120, 833

\bibitem[{{Rich} {et~al.}(2017){Rich}, {Ryde}, {Thorsbro}, {Fritz}, {Schultheis}, {Origlia}, \& {J{\"o}nsson}}]{rich:17}
{Rich}, R.~M., {Ryde}, N., {Thorsbro}, B., {et~al.} 2017, {\it AJ}, 154, 239

\bibitem[{{Ryde} {et~al.}(2025){Ryde}, {Nandakumar}, {Schultheis}, {Kordopatis}, {di Matteo}, {Haywood}, {Sch{\"o}del}, {Nogueras-Lara}, {Rich}, {Thorsbro}, {Mace}, {Agertz}, {Amarsi}, {Kocher}, {Molero}, {Orglia}, {Pagnini}, \& {Spitoni}}]{ryde:2025}
{Ryde}, N., {Nandakumar}, G., {Schultheis}, M., {et~al.} 2025, \apj, 979, 174

\bibitem[{{Sch{\"o}del} {et~al.}(2014){Sch{\"o}del}, {Feldmeier}, {Neumayer}, {Meyer}, \& {Yelda}}]{schodel_nsc:2014}
{Sch{\"o}del}, R., {Feldmeier}, A., {Neumayer}, N., {Meyer}, L., \& {Yelda}, S. 2014, Classical and Quantum Gravity, 31, 244007

\bibitem[{{Sch\"odel} {et~al.}(2023){Sch\"odel}, {Longmore}, {Henshaw}, {Ginsburg}, {Bally}, {Feldmeier}, {Hosek}, {Nogueras Lara}, {Ciurlo}, {Chevance}, {Kruijssen}, {Klessen}, {Ponti}, {Amaro-Seoane}, {Anastasopoulou}, {Anderson}, {Arias}, {Barnes}, {Battersby}, {Bono}, {Bravo Ferres}, {Bryant}, {Cano Gonz{\'a}alez}, {Cassisi}, {Chaves-Velasquez}, {Conte}, {Contreras Ramos}, {Cotera}, {Crowe}, {di Teodoro}, {Do}, {Eisenhauer}, {Enokiya}, {Fedriani}, {Friske}, {Gadotti}, {Gallart}, {Gallego Calvente}, {Gallego Cano}, {Garc{\'\i}a Fuentes}, {Garc{\'\i}a Mar{\'\i}n}, {Gardini}, {Gautam}, {Ghez}, {Gillessen}, {Gouda}, {Gualandris}, {Guarcello}, {Gutermuth}, {Haggard}, {Hankins}, {Hu}, {Kano}, {Kauffmann}, {Lau}, {Lazarian}, {Libralato}, {Lu}, {Lu}, {Lu}, {Luetzgendorf}, {Magorrian}, {Mandel}, {Markoff}, {Mart{\'\i}nez Arranz}, {Mastrobuono-Battisti}, {Melamed}, {Mills}, {Mori}, {Morris}, {Murchikova}, {Nagata}, {Najarro}, {Nandakumar}, {Nataf}, {Neumayer}, {Nishiyama}, {Nobukawa}, {Par{\'e}}, {Peissker},
  {Petkova}, {Pillai}, {Rom{\'a}n}, {Rugel}, {Ryde}, {Sabha}, {S{\'a}nchez Berm{\'u}dez}, {S{\'a}nchez-Monge}, {Schultheis}, {Shao}, {Shinnaga}, {Simpson}, {Takekawa}, {Tan}, {Thorsbro}, {Torne}, {Goppala Tress}, {Uchiyam}, {Valenti}, {van der Marel}, {Verberne}, {Vermot}, {von Fellenberg}, {Walker}, {Witzel}, {Xu}, {Yano}, {Yusef-Zadeh}, {Zaja{\v{c}}ek}, \& {Zoccali}}]{jwst_nsc_white:23}
{Sch\"odel}, R., {Longmore}, S., {Henshaw}, J., {et~al.} 2023, arXiv e-prints, arXiv:2310.11912

\bibitem[{{Sch{\"o}del} {et~al.}(2020){Sch{\"o}del}, {Nogueras-Lara}, {Gallego-Cano}, {Shahzamanian}, {Gallego-Calvente}, \& {Gardini}}]{schodel:2020}
{Sch{\"o}del}, R., {Nogueras-Lara}, F., {Gallego-Cano}, E., {et~al.} 2020, \aap, 641, A102

\bibitem[{{Skrutskie} {et~al.}(2006){Skrutskie}, {Cutri}, {Stiening}, {Weinberg}, {Schneider}, {Carpenter}, {Beichman}, {Capps}, {Chester}, {Elias}, {Huchra}, {Liebert}, {Lonsdale}, {Monet}, {Price}, {Seitzer}, {Jarrett}, {Kirkpatrick}, {Gizis}, {Howard}, {Evans}, {Fowler}, {Fullmer}, {Hurt}, {Light}, {Kopan}, {Marsh}, {McCallon}, {Tam}, {Van Dyk}, \& {Wheelock}}]{2mass}
{Skrutskie}, M.~F., {Cutri}, R.~M., {Stiening}, R., {et~al.} 2006, {\it AJ}, 131, 1163

\bibitem[{{Sneden} {et~al.}(2014){Sneden}, {Lucatello}, {Ram}, {Brooke}, \& {Bernath}}]{sneden:14}
{Sneden}, C., {Lucatello}, S., {Ram}, R.~S., {Brooke}, J.~S.~A., \& {Bernath}, P. 2014, \apjs, 214, 26

\bibitem[{{Solanki} {et~al.}(2023){Solanki}, {Ressler}, {Murchikova}, {Stone}, \& {Morris}}]{solanki:2023}
{Solanki}, S., {Ressler}, S.~M., {Murchikova}, L., {Stone}, J.~M., \& {Morris}, M.~R. 2023, \apj, 953, 22

\bibitem[{{Sormani} {et~al.}(2022){Sormani}, {Sanders}, {Fritz}, {Smith}, {Gerhard}, {Sch{\"o}del}, {Magorrian}, {Neumayer}, {Nogueras-Lara}, {Feldmeier-Krause}, {Mastrobuono-Battisti}, {Schultheis}, {Shahzamanian}, {Vasiliev}, {Klessen}, {Lucas}, \& {Minniti}}]{sormani:2022}
{Sormani}, M.~C., {Sanders}, J.~L., {Fritz}, T.~K., {et~al.} 2022, \mnras, 512, 1857

\bibitem[{{Sormani} {et~al.}(2020){Sormani}, {Tress}, {Glover}, {Klessen}, {Battersby}, {Clark}, {Hatchfield}, \& {Smith}}]{Sormani:2020}
{Sormani}, M.~C., {Tress}, R.~G., {Glover}, S. C.~O., {et~al.} 2020, \mnras, 497, 5024

\bibitem[{{Taniguchi} {et~al.}(2025){Taniguchi}, {Matsunaga}, {Kobayashi}, {Jian}, {Thorsbro}, {Fukue}, {Hamano}, {Ikeda}, {Kawakita}, {Kondo}, {Otsubo}, {Sameshima}, {Tsujimoto}, \& {Yasui}}]{taniguchi:2025}
{Taniguchi}, D., {Matsunaga}, N., {Kobayashi}, N., {et~al.} 2025, \aap, 693, A163

\bibitem[{{Thorsbro}(2020)}]{thorsbro_atoms:20}
{Thorsbro}, B. 2020, Atoms, 8, 4

\bibitem[{{Thorsbro} {et~al.}(2023){Thorsbro}, {Forsberg}, {Kordopatis}, {Mastrobuono-Battisti}, {Church}, {Rich}, {Ryde}, {Schultheis}, \& {Nishiyama}}]{thorsbro:2023}
{Thorsbro}, B., {Forsberg}, R., {Kordopatis}, G., {et~al.} 2023, \apjl, 958, L18

\bibitem[{Thorsbro {et~al.}(2017)Thorsbro, Ryde, Rich, Schultheis, Fritz, \& Origlia}]{thorsbro:17}
Thorsbro, B., Ryde, N., Rich, R.~M., {et~al.} 2017, Proceedings of the International Astronomical Union, 13, 372–373

\bibitem[{{Thorsbro} {et~al.}(2020){Thorsbro}, {Ryde}, {Rich}, {Schultheis}, {Renaud}, {Spitoni}, {Fritz}, {Mastrobuono-Battisti}, {Origlia}, {Matteucci}, \& {Sch{\"o}del}}]{thorsbro:2020}
{Thorsbro}, B., {Ryde}, N., {Rich}, R.~M., {et~al.} 2020, \apj, 894, 26

\bibitem[{{Thorsbro} {et~al.}(2018){Thorsbro}, {Ryde}, {Schultheis}, {Hartman}, {Rich}, {Lomaeva}, {Origlia}, \& {J{\"o}nsson}}]{thorsbro:2018}
{Thorsbro}, B., {Ryde}, N., {Schultheis}, M., {et~al.} 2018, \apj, 866, 52

\bibitem[{{Ting} {et~al.}(2019){Ting}, {Conroy}, {Rix}, \& {Cargile}}]{ting:2019}
{Ting}, Y.-S., {Conroy}, C., {Rix}, H.-W., \& {Cargile}, P. 2019, \apj, 879, 69

\bibitem[{{Tody}(1993)}]{IRAF}
{Tody}, D. 1993, in ASP Conf. Ser. 52: Astronomical Data Analysis Software and Systems II, ed. R.~J. {Hanisch}, R.~J.~V. {Brissenden}, \& J.~{Barnes}, 173

\bibitem[{{Tremaine} {et~al.}(1975){Tremaine}, {Ostriker}, \& {Spitzer}}]{Tremaine75}
{Tremaine}, S.~D., {Ostriker}, J.~P., \& {Spitzer}, Jr., L. 1975, \apj, 196, 407

\bibitem[{{Tsatsi} {et~al.}(2017){Tsatsi}, {Mastrobuono-Battisti}, {van de Ven}, {Perets}, {Bianchini}, \& {Neumayer}}]{Tsatsi:2017}
{Tsatsi}, A., {Mastrobuono-Battisti}, A., {van de Ven}, G., {et~al.} 2017, \mnras, 464, 3720

\bibitem[{{Valenti} \& {Piskunov}(2012)}]{sme_code}
{Valenti}, J.~A. \& {Piskunov}, N. 2012, {SME: Spectroscopy Made Easy}, astrophysics Source Code Library

\end{thebibliography}

\end{document}